# A new relation between the hot electron power loss and acoustic phonon limited mobility in Bloch- Grüneisen regime

S. S. Kubakaddi

Department of Physics, K. L. E. Technological University, Hubballi-580031, Karnataka, India
Email: sskubakaddi@gmail.com
(June 20, 2017)

Expressions for the electron power loss $F(T)$ and mobility $\mu_p$ due to acoustic phonon scattering are given in the Bloch- Grüneisen (BG) regime for three- and two- dimensional electron gas in semiconductors and Dirac-fermions. We obtain a simple relation $F(T)\mu_p = \eta e v_s^2$, where $\eta$ (~1) is a constant, $e$ is the electron charge and $v_s$ is the acoustic phonon velocity. It is found to be independent of temperature and electron concentration. This relation is applied to GaAs heterojucntions and graphene, to obtain $\mu_p$ from the measured $F(T)$. We propose that, using this relation, the measurements of $F(T)$, in BG regime, which depends exclusively upon the electron - acoustic phonons coupling, could serve as a tool to determine the low temperature $\mu_p$, which is otherwise difficult to measure due to the contributions from the lattice disorders.

## I. Introduction

The scattering mechanisms that govern the transport coefficients in semiconductors are lattice disorders (impurities and defects) and lattice vibrations (phonons) [1]. In disorder free systems or in the clean limit, phonons are the only scattering sources. Hence, phonon scattering, governed by electron-phonon (el-ph) coupling strength, is supposed to set the fundamental limit on the transport coefficients such as electrical conductivity and mobility and there by on the electronic quality and possible performance of the devices. Mobility is one of the most important figures of merit for any electronic material. It is dominated by the acoustic (optical) phonon scattering at low (high) temperature. For e. g. acoustic phonon limited mobility $\mu_p$ is found to be ~$10^9$ cm$^2$/ V-s, at about 1K, in conventional two-dimensional electron gas (2DEG) in GaAs heterojunctions (HJs) [2] and $\mu_p$ >$2\times10^5$cm$^2$/V-s at room temperature in graphene [3,4].

Phonon-drag thermopower $S^g$, arising due to the dragging of electrons by the phonon wind in the temperature gradient, is also purely governed by the electron-acoustic phonon coupling and dominates the thermopower at low temperature (< about 10 K). The basic scattering mechanism involved in $S^g$ and acoustic phonon limited mobility $\mu_p$ is only the electron-acoustic phonon interaction. At very low temperature, i.e. in the Bloch- Grüneisen (BG) regime ($q << 2k_f$, where $q$ ($k_f$) is the phonon (Fermi) wave vector), an important relation between $S^g$ and $\mu_p$ is given by Herring's law $S^g\mu_p= (f\Lambda v_s)T^{-1}$ [5], where f is a fraction and $\Lambda$ is the phonon mean free path, $v_s$ is the velocity of phonons and $T$ is the lattice temperature. It is shown to be valid in three- dimensional (3D) and two-dimensional (2D) semiconductors [5-7] and Dirac fermions [8,9]. Importance of this relation is that, if one of them is measured, then the other one can be found. At low $T$, it is difficult to extract and determine experimentally $\mu_p$ because of the contribution from the lattice disorders and uncertainty in their scattering strength. Hence, from the $S^g$ measurements $\mu_p$ can be determined applying Herring's law. For example, the $S^g$ of 2DEG at zero magnetic field and composite fermions (CF's) at high magnetic fields fields in GaAs/Ga$_{1-x}$Al$_x$As HJs, measured in the temperature range 0.1–1.2 K, have been used to evaluate the acoustic-phonon limited mobility of electrons and CF's as a function of temperature [10].

Hot electron power loss $P$, is another property governed exclusively by el-ph coupling. In large electric field, electrons establish their own 'hot electron temperature' $T_e > T$ and lose their energy by emission of phonons. At low temperature, emission of acoustic phonons is the only channel through which hot electron energy relaxation and cooling takes place. This has been investigated extensively in the 3D semiconductor [11] and 2DEG in GaAS HJs [12-14] and Si-MOSFETs [15], and Dirac fermions in graphene [8,16-18], 3D Dirac semimetals [19,20] and transition metal dichalcogenides (TMDs) [21]. It was shown that, in BG regime, $P$ and $S^g$ are related by $P=\xi e v_s T S^g/\Lambda$, where $\xi$ (~1) is a numerical constant and $e$ is the electron charge [8,15,20].

Both $P$ and $\mu_p$ are the sensitive measures of electron-acoustic phonon coupling but in different ways. $P$ is determined by the energy relaxation whereas $\mu_p$ involves momentum relaxation through electron-acoustic phonon interaction. Since the underlying mechanism for both the properties is same, we expect a relation between these two properties. In the present work, in the Bloch- Grüneisen regime, we find a relation between $P$ and $\mu_p$. We obtain the product $P\mu_p$ for electrons in 3D and 2D semiconductors and Dirac fermions. In order to arrive at this relation, we briefly give how $P$ and $\tau_p$, the momentum relaxation time, and hence $\mu_p$ due to acoustic phonon scattering are obtained. We discuss the dependence of the product $P\mu_p$ on various parameters. We numerically find $\mu_p$ from measured $P$ in BG regime in GaAs HJs and graphene monolayer.

## II. Basic equations

The hot electron power loss $P$, per electron, due to electron interaction with the acoustic phonons of frequency $\omega_\mathbf{q}$ and wave vector $\mathbf{q}$ is given by [13,22]

$$P = \frac{1}{N_e}\sum_\mathbf{q}\hbar\omega_\mathbf{q}\left(\frac{dN_\mathbf{q}}{dt}\right), \tag{1}$$

where $N_e$ is the total number of electrons and the phonon rate equation is

$$\left(\frac{dN_\mathbf{q}}{dt}\right) = \left(\frac{2\pi}{\hbar}\right)\sum_\mathbf{k}|C(q)|^2\{N_\mathbf{q}(T)f(E_\mathbf{k})[1-f(E_{\mathbf{k'}})] \\ -(N_\mathbf{q}(T)+1)f(E_{\mathbf{k'}})[1-f(E_\mathbf{k})]\}\delta(E_{\mathbf{k'}}-E_\mathbf{k}-\hbar\omega_\mathbf{q})\delta_{\mathbf{k'},\mathbf{k}+\mathbf{q}}. \tag{2}$$

In the above equation $|C(q)|^2$ is the square of the electron - acoustic phonon matrix element, $N_q$ is the Bose distribution for phonons at lattice temperature $T$, $E_k$ is the energy of the electron in state $\mathbf{k}$ and $f(E_k)$ is the Fermi-Dirac distribution at electron temperature $T_e$ with the Fermi energy $E_f$. The above power loss equation can be expressed as [14]

$$P = F(T_e) - F(T), \quad (3)$$

where

$$F(T) = \frac{1}{N_e}\left(\frac{2\pi}{\hbar}\right)\sum_{\mathbf{k'}}\hbar\omega_\mathbf{q}\sum_{\mathbf{k}}|C(q)|^2 N_\mathbf{q}(T) \quad (4)$$
$$\times [f(E_\mathbf{k} + \hbar\omega_\mathbf{q}) - f(E_\mathbf{k'})]\delta(E_\mathbf{k'} - E_\mathbf{k} - \hbar\omega_\mathbf{q})\delta_{\mathbf{k'},\mathbf{k}+\mathbf{q}}.$$

For fermions, the equations for $F(T)$, in different systems, are obtained with the parabolic energy dispersion $E_k = (\hbar k)^2/2m$ ($m$ being the effective mass of the electron) and the density of states $D(E_k) = (gm/2\pi\hbar^2)$ for 2D and $(g/\sqrt{2}\pi)(m/\hbar)^{3/2}E_k^{1/2}$ for 3D electron gas. In case of Dirac-fermions the energy dispersion is linear $E_\mathbf{k} = \hbar v_f k$ with the density of states $D(E_\mathbf{k}) = gE_\mathbf{k}/(2\pi\hbar^2 v_f^2)$ [3] and $gE_\mathbf{k}^2/[2\pi^2(\hbar v_f)^3]$ [23], respectively, for 2D and 3D systems. Here, $g = (g_s g_v)$ is the electron (spin and valley) degeneracy and $v_f$ is the Fermi velocity.

Interestingly, we find that for $T_e >> T$ or $T$ close to zero, $P = F(T_e)$. The equations for $F(T)$, obtained in the Bloch-Grüneisen regime, for 3D and 2D semiconductors and the Dirac fermions are given in Table I. The $F(T)$ for 2DEG in transition metal dichalcogenides [21], 3D Dirac fermions [20] and 2D Dirac fermions [8] are obtained by us. The temperature $T$ and concentration $n$ dependence of the power loss are, respectively, given by the power laws $F(T) \sim T^r$ and $n^{-s}$, where $r$ and $s$ are the positive constant exponents.

The energy dependent momentum relaxation time due to electron scattering by acoustic phonons is defined by [23,24]

$$(1/\tau(E_\mathbf{k})) = \sum_{\mathbf{k'}}(1-\cos\theta)[1-f(E_{\mathbf{k'}})]/[1-f(E_\mathbf{k})] \quad (5)$$

where $\theta$ is the angle between $\mathbf{k}$ and $\mathbf{k'}$ and

$$W_{\mathbf{k}\mathbf{k'}} = \frac{2\pi}{\hbar}|C(q)|^2[N_\mathbf{q}\delta(E_\mathbf{k} - E_{\mathbf{k'}} + \hbar\omega_\mathbf{q})\delta_{\mathbf{k'},\mathbf{k}+\mathbf{q}} \quad (6)$$
$$+ (N_\mathbf{q}+1)\delta(E_\mathbf{k} - E_{\mathbf{k'}} - \hbar\omega_\mathbf{q})\delta_{\mathbf{k'},\mathbf{k}-\mathbf{q}}]$$

is the transition probability from state $\mathbf{k}$ to $\mathbf{k'}$.

The relaxation time $\tau(E_f)$, thus obtained, at $E_\mathbf{k} = E_f$ is used to find the acoustic phonon limited mobility $\mu_p$ in the BG regime. In semiconductors $\mu_p = e\tau(E_f)/m$. For Dirac fermions $\mu_p = \sigma/ne$, where $\sigma = e^2 v_f^2 D(E_f)\tau(E_f)/2$ for 2D Dirac fermions [3] and $\sigma = e^2 v_f^2 D(E_f)\tau(E_f)/3$ for 3D Dirac fermions [23]. The $\mu_p$, thus obtained, are listed in Table I. The $\mu_p$ for 2DEG in TMDs and 3D Dirac fermions are obtained by us in the present work. These expressions can be represented by the power laws $\mu_p \sim T^{-r}$ and $n^s$.

Taking the product of $F(T)$ and $\mu_p$, in all the systems listed in the Table I, we find a simple relation

$$F(T)\mu_p = \eta e v_s^2, \quad (7)$$

where $\eta$ (~1) is a constant and $v_s$ is the velocity of acoustic phonons. In the equations of Table I, $D$ is the acoustic deformation potential constant, $h_{14}$ is the piezoelectric constant, $n_v$ ($n_s$) is the 3D (2D) electron concentration, $\rho_s(\rho_v)$ is the surface (volume) mass density, $v_{sl}$ ($v_{st}$) is the velocity of longitudinal (transverse) acoustic phonons and $\zeta(n)$ is the Riemann zeta function. We expect the relation (7) to be valid when screening of el-ph interaction is taken in to account, as the screening affects both the energy and momentum relaxation in the same way.

The product $F(T) \mu_p$ exclusively depends on only the velocity of acoustic phonons in the respective systems. Interestingly, this product is found to be independent of temperature and electron concentration and other material parameters. In Table I, we notice that $0 < \eta < 1.5$, and it is found to be different for different systems and different coupling mechanisms. The difference in $\eta$ values for different electron systems may be attributed to the dimensionality of phonons and electrons and the electron dispersion.

The equations given for 2DEG in HJs and monolayer graphene are in the clean limit. For graphene on the substrate there is no contribution from the flexural phonons [26-27]. Contribution to $F(T)$ from the vector potential coupling has the same $T$ and $n_s$ dependence as that of unscreened deformation potential coupling [27]. In graphene, the experiments measuring hot electron power loss [17,18] and their compatibility with the theoretical predictions for the unscreened deformation potential coupling [8] suggests that screening does not seem to be playing the role.

### III. Discussion

In Table I, the power laws with regard to the temperature dependence are characteristic of dimensionality of the phonons (with linear dispersion) and they are independent of dimensionality of the electron system. Whereas, the power laws with regard to electron concentration are determined by the dimensionality and dispersion of the electron system. However, in a given system, the nature of electron dispersion determines the power law for the electron concentration.

Since, for $T_e >> T$ or $T$ close to zero, $P = F(T_e)$, the measurements of $F(T)$ under such conditions can be used to determine $\mu_p$.

In BG regime, in 3D polar semiconductors and for 2DEG in GaAs HJs, scattering due to the piezoelectric coupling is expected to dominate $F(T)$ and $\mu_p$ due to $q^{-1}$ dependence of its matrix element as compared to $q$ dependence of the matrix element of deformation potential coupling [2,11,14]. Where as, for 2DEG in Si-MOFETs [15] and 2DHG in SiGe HJs [28] the low temperature phonon limited transport is governed by the scattering only due to acoustic phonon deformation potential coupling.

We give an estimate of $\mu_p$ for 2DEG in a GaAs HJ and graphene from the measured $F(T)$. In GaAs HJ, at low temperatures, the electron scattering is due to both the deformation potential and piezoelectric coupling. For sub-Kelvin temperatures and up to 1 K piezoelectric scattering is dominating [2,10,14]. We take the $F(T)$ data from Ma et al [14]. In Fig. (7) of Ref. [14] the observations are for zero lattice temperature and we read $F(T) \approx 20$ eV/s at $T = 1$ K. Taking $v_{st} = 3.01 \times 10^5$ cm/s, using relation (7), we get, $\mu_p = 6.034 \times 10^9$ cm$^2$/V-s which is of the order of experimentally observed and theoretically calculated $\mu_p$ by Stormer et al [2]. It is to be recalled that $\mu_p$ is obtained in the experiment of Ref.[2] by subtracting $T$ - independent contribution due to impurity scattering, using Matthiessen's rule.

In graphene, at low temperature, electron scattering by the acoustic phonons is only due to the deformation potential coupling [3]. For illustration in graphene, we take the

**Table I**: The expressions and power laws for the electron power loss $F(T)$ and mobility $\mu_p$ due to acoustic phonon coupling in the Bloch-Grüneisen regime for different electron systems.

| Electron system | Power loss $F(T)$ | Acoustic phonon limited mobility $\mu_p$ | $\eta$ |
|---|---|---|---|
| 3D semiconductor (3D fermions): Deformation potential coupling | $\dfrac{6D^2 m^2 (k_B T)^5 \zeta(5)}{\pi^3 \rho_v \hbar^7 v_{sl}^4 n_v}$ [11]* | $\dfrac{\pi^2 e \rho_v \hbar^7 v_{sl}^6 n_v}{10 D^2 m^2 (k_B T)^5 \zeta(5)}$ (a) | 0.191 |
| 3D semiconductor (3D fermions): Piezoelectric coupling | $\dfrac{(ee_{14})^2 m^2 (k_B T)^3}{2\pi^3 \rho_v \hbar^5 \varepsilon^2 v_{st}^2 n_v}\left(\dfrac{16}{35}+\dfrac{12 v_{st}^2}{35 v_{sl}^2}\right)$ [11]* | $\dfrac{2\pi^2 e \rho_v \hbar^5 \varepsilon^2 v_{st}^4 n_v}{(ee_{14})^2 m^2 (k_B T)^3}\left(\dfrac{16}{35}+\dfrac{12 v_{st}^2}{35 v_{sl}^2}\right)^{-1}$ (a) | $1/\pi$ |
| 3D Dirac semimetal (3D Dirac-fermions): Deformation potential coupling | $\dfrac{3 g^{1/3} D^2 (k_B T)^5 4! \zeta(5)}{4\pi^2 (6\pi)^{1/3} \rho_v \hbar^5 v_f^2 v_{sl}^4 n_v^{1/3}}$ [20] | $\dfrac{4(6\pi)^{1/3} e \rho_v \hbar^5 v_{sl}^6 v_f^2 n_v^{1/3}}{g^{1/3} D^2 (k_B T)^5 5! \zeta(5)}$ (b) | 0.061 |
| 2DEG in semiconductor HJs (2D fermions): Deformation potential coupling | $\dfrac{12 D^2 m^2 (k_B T)^5 \zeta(5)}{2^{3/2} \pi^{5/2} \rho_s \hbar^7 v_{sl}^4 n_s^{3/2}}$ [14] | $\dfrac{2^{3/2} \pi^{5/2} e \rho_s \hbar^7 v_{sl}^6 n_s^{3/2}}{15 D^2 m^2 (k_B T)^5 \zeta(5)}$ [25] | 0.8 |
| 2DEG in semiconductor HJs (2D fermions): Piezoelectric coupling | $\dfrac{(eh_{14})^2 m^2 (k_B T)^3 \zeta(3)}{64\sqrt{2} \pi^{5/2} \rho_s \hbar^5 n_s^{3/2} v_{st}^2}(13+9\gamma^2)$ [14] | $\dfrac{(2\pi)^{5/2} e \rho_v \hbar^5 \varepsilon^2 v_{st}^4 n_s^{3/2}}{(eh_{14})^2 m^2 (k_B T)^3 \zeta(3)} \dfrac{256}{177}\left(1+\dfrac{45\gamma^4}{59}\right)^{-1}$ [25] | 1.332 for $\gamma=0.59$ [14,25] |
| Monolayer of transition metal dichalcogenides (2D fermions): Deformation potential coupling (e) | $\dfrac{\pi^{3/2} D^2 m^2 (k_B T)^4}{15 \hbar^6 \rho_s v_{sl}^3 n_s^{3/2}}$ [21] | $\dfrac{2\pi^{5/2} e \rho_s \hbar^6 v_{sl}^5 n_s^{3/2}}{D^2 m^2 (k_B T)^4 4! \zeta(4)}$ (c) | 0.5 |
| Monolayer of graphene (2D Dirac-fermions): Deformation potential coupling | $\dfrac{D^2 (k_B T)^4 3! \zeta(4)}{\pi^{3/2} \rho_s \hbar^4 v_f^3 v_{sl}^2 n_s^{1/2}}$ [8] | $\dfrac{2\pi^{3/2} e \rho_s \hbar^4 v_{sl}^5 v_f^2 n_s^{1/2}}{D^2 (k_B T)^4 4! \zeta(4)}$ (d) | 0.5 |

* In Eq.(8.27) of Ref.[11], in the denominator, the electron concentration $n_v$ seems to be missing.
(a) $\mu_p$ is obtained using $\tau_p(E_f)$ from Ref. [11].
(b) and (c): $\mu_p$ are obtained by us. We note that the equation for $\tau_f^{-1}$ obtained by us for 3DDirac semimetal is differing from the one obtained in Ref. [23]. Factor 8 in the numerator of Ref. [23] is to be replaced by 1/2.
(d) $\mu_p$ is obtained using $\tau_p(E_f)$ from Ref. [3].
(e) We also find that, in bilayer graphene the BG regime expressions for $F(T)$ and $\mu_p$ are similar to as found for monolayer TMDs.

$F(T)$ data from Fig. 7 of Baker et al [18] for the sample with $n_s=1.62\times10^{13}$ cm$^{-2}$. We read, approximate experimental $F(T)$ to be about $1.5\times10^{-16}$, $7\times10^{-16}$ and $5\times10^{-13}$ W, respectively at 2, 3 and 20 K. Using the value of $v_{sl}=2\times10^6$ cm/s, we obtain $\mu_p = 2.14\times10^9$, $4.6\times10^8$ and $6.41\times10^5$ cm$^2$/V-s, respectively, at 2, 3 and 20K. In graphene, the low temperature experimental data of acoustic phonon limited resistivity $\rho_p$ is obtained by subtracting the residual resistivity, that stems from almost the temperature independent electron scattering by the static impurities and point defects [29]. Thus obtained $\rho_p \sim T^4$ behaviour is a characteristic of scattering by 2D phonons in BG regime. Taking approximately $\rho_p = 0.2\Omega$ at 20 K, from the data of Ref. [29], for the sample with $n_s=1.36\times10^{13}$ cm$^{-2}$, the corresponding $\mu_p = 2.3\times10^6$ cm$^2$/V-s and from the relation (7), $F(T) = 1.4\times10^{-13}$ W. This $F(T)$ is nearly 3.5 times smaller than the $F(T)$ observed by Baker et al [18]. It is to be noted that the electron concentration $n_s$ in the sample of Baker et al [18] and Efetov et al [29] are marginally differing. Normalising the value of $F(T)$ obtained from the data of Efetov et al to the $n_s$ value of Baker et al, knowing $F(T) \sim n_s^{-1/2}$, we get $F(T) = 1.26\times10^{-13}$ W, which is about four times smaller than the value observed for Baker et al sample. However, there is a need for the low T (< 10 K) mobility data where BG regime is strictly valid.

With the above illustrations, we emphasize that the relation between $F(T)$ and $\mu_p$ can be used as a new tool for determining $\mu_p$ from the measured $F(T)$. Inturn, the electron-acoustic phonon coupling constants $D$ and $h_{14}$ can also be determined unambiguously.

**IV. Conclusions**

In conclusion, we have obtained a new relation between the hot electron power loss and phonon limited mobility in the Bloch-Grüneisen regime. It is found to depend only upon the acoustic phonon velocity and is independent of temperature, electron concentration and other material parameters. This relation can be used as a new tool to find acoustic phonon limited mobility in BG regime, analogous to Herring's law [5], by measuring hot electron power loss. Our predicted relation has an advantage over the Herring's law in the cases where phonon-drag is not significant enough to measure compared to the diffusion component of thermopower.